\documentclass{osa-article}

\journal{osajournal}

\articletype{Research Article}

\usepackage{amssymb}

\begin{document}

\title{Progress in the measurement and reduction of thermal noise in optical coatings for gravitational-wave detectors}

\author{M. Granata,\authormark{1,*} A. Amato,\authormark{1}, G. Cagnoli\authormark{2}, M. Coulon,\authormark{1}, J. Degallaix,\authormark{1}, D. Forest,\authormark{1}, L. Mereni,\authormark{1}, C. Michel,\authormark{1}, L. Pinard,\authormark{1}, B. Sassolas,\authormark{1} J. Teillon\authormark{1}}

\address{\authormark{1}Laboratoire des Mat\'{e}riaux Avanc\'{e}s - IP2I, CNRS, Universit\'{e} de Lyon, F-69622 Villeurbanne, France\\
\authormark{2}Universit\'{e} de Lyon, Universit\'{e} Claude Bernard Lyon 1, CNRS, Institut Lumi\`{e}re Mati\`{e}re, F-69622 Villeurbanne, France\\}

\email{\authormark{*}m.granata@lma.in2p3.fr}

\begin{abstract}
Coating thermal noise is a fundamental limit for precision experiments based on optical and quantum transducers. In this review, after a brief overview of the techniques for coating thermal noise measurements, we present the latest world-wide research activity on low-noise coatings, with a focus on the results obtained at the Laboratoire des Mat\'{e}riaux Avanc\'{e}s. We report new updated values for the Ta$_2$O$_5$, Ta$_2$O$_5$-TiO$_2$ and SiO$_2$ coatings of the Advanced LIGO, Advanced Virgo and KAGRA detectors, and new results from sputtered Nb$_2$O$_5$, TiO$_2$-Nb$_2$O$_5$, Ta$_2$O$_5$-ZrO$_2$, MgF$_2$, AlF$_3$ and silicon nitride coatings. Amorphous silicon, crystalline coatings, high-temperature deposition, multi-material coatings and composite layers are also briefly discussed, together with the latest developments of structural analyses and models.
\end{abstract}

\section{The issue of coating thermal noise}
The {\it Laboratoire des Mat\'{e}riaux Avanc\'{e}s} (LMA, now a division of the newly created {\it Institut de Physique des 2 Infinis de Lyon}) has provided the current high-reflection (HR) and anti-reflective (AR) coatings of the most critical optics of Advanced LIGO \cite{Aasi15}, Advanced Virgo \cite{Acernese15} and KAGRA \cite{Aso13}. In these gravitational-wave (GW) interferometers, large and massive suspended mirrors (typically with $\varnothing$ = 35 cm, t = 20 cm, m = 40 kg) form the km-long resonant Fabry-Perot cavities where the astrophysical signals are embedded in the laser beam phase. Their HR coatings are Bragg reflectors of alternate layers of ion-beam-sputtered (IBS) low- and high-refractive index materials, which feature outstanding optical properties \cite{Pinard17}. At the same time, these amorphous coatings are the source of coating thermal noise (CTN), which is a severe limitation to the detector sensitivity \cite{Aasi15,Acernese15}.

In GW interferometers, thermal noise arises from fluctuations of the mirror surface under the random motion of particles in coatings and substrates \cite{Saulson90,Levin98}. Its power spectral density is determined by the amount of internal friction within the mirror materials, via the fluctuation-dissipation theorem \cite{Callen52}: the higher the elastic energy loss, the higher the thermal noise level. As the coating loss is usually several orders of magnitude larger than that of the substrate \cite{Crooks02,Harry02}, CTN is the dominant source of noise in the mirrors.

More generally, CTN is a fundamental limit for precision experiments based on optical and quantum transducers, such as opto-mechanical resonators \cite{Aspelmayer14}, frequency standards \cite{Matei17} and quantum computers \cite{Martinis05}. In the last two decades, a considerable research effort has been committed to the measurement and the reduction of CTN.

\section{Measurement}
Besides km-scale interferometry, several techniques are available to measure CTN. The first direct approach has been to measure the differential displacement of two suspended identical Fabry-Perot cavities in a table-top setup \cite{Numata03,Black04}. More recently, two alternative single-cavity experiments have been successfully achieved: a single-crystal silicon Fabry-Perot cavity for cryogenic operation \cite{Kessler12}, and a folded Fabry-Perot cavity operated with higher-order laser modes \cite{Gras17,Gras18}; the latter is the current reference solution for CTN measurements of the LIGO Scientific Collaboration.

An alternative option for direct CTN measurements is quadrature phase differential interferometry \cite{Bellon02}, where polarized laser beams are used in a modified Nomarski interferometer with complex contrast to measure the noise-driven displacement of a coated atomic-force microscope silicon tip \cite{Li14}.

\begin{figure}
	\centering\includegraphics[width=\textwidth]{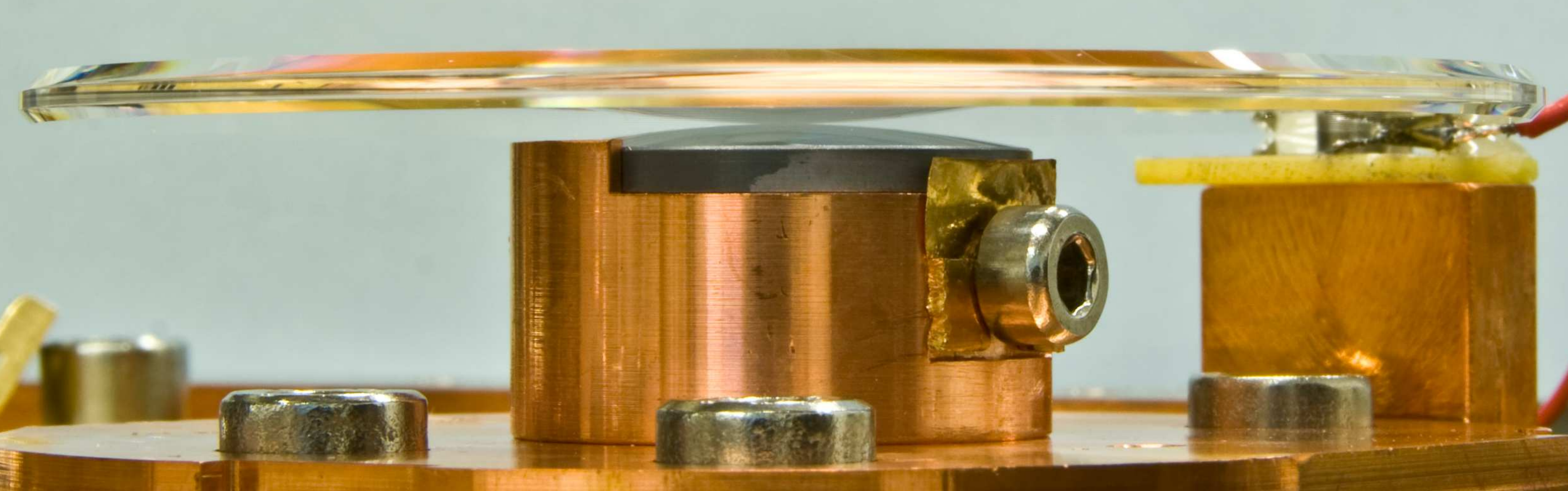}
	\caption{\label{FIG_GeNS}Coated fused-silica disk ($\varnothing =3$") on top of a cryogenic GeNS system, composed of a silicon plano-convex lens in a copper mount. An electrostatic drive, used for contactless excitation of the normal modes of the disk, is visible on the right (in the background).}
\end{figure}

An indirect approach is the measurement of the coating internal friction $\phi_c$ in a suspended coated resonator \cite{Berry75}, though such measurement could be easily degraded by systematics due to the suspension system. At the LMA, we opted for a Gentle Nodal Suspension (GeNS) system \cite{Cesarini09} and adapted it for cryogenic measurements as well \cite{Granata15}. In such a system, shown on Fig. \ref{FIG_GeNS}, a disk-shaped resonator of thickness $t$ is installed on top of a spherical surface of radius $r$; as long as $t < 2r$ and there is no sliding between the contact surfaces, the disk is in stable equilibrium. The system can accept disks of different diameters: so far, we tested samples ranging from 1" to 3". The key features of the GeNS system are: i) absence of clamping; ii) extremely low excess losses ($\phi < 5\cdot10^{-9}$ measured for 3" samples) coming from the point-like suspension surface; iii) unprecedented reproducibility of measurements, within few percent on internal friction and 0.01\% on resonant frequencies; iv) the possibility to use easily available substrates (like silicon wafers, for instance). For all these reasons, the GeNS system is now a standard setup within the Virgo and LIGO Collaborations \cite{Granata16,Vajente17}.

\begin{figure}
	\centering\includegraphics[width=\textwidth]{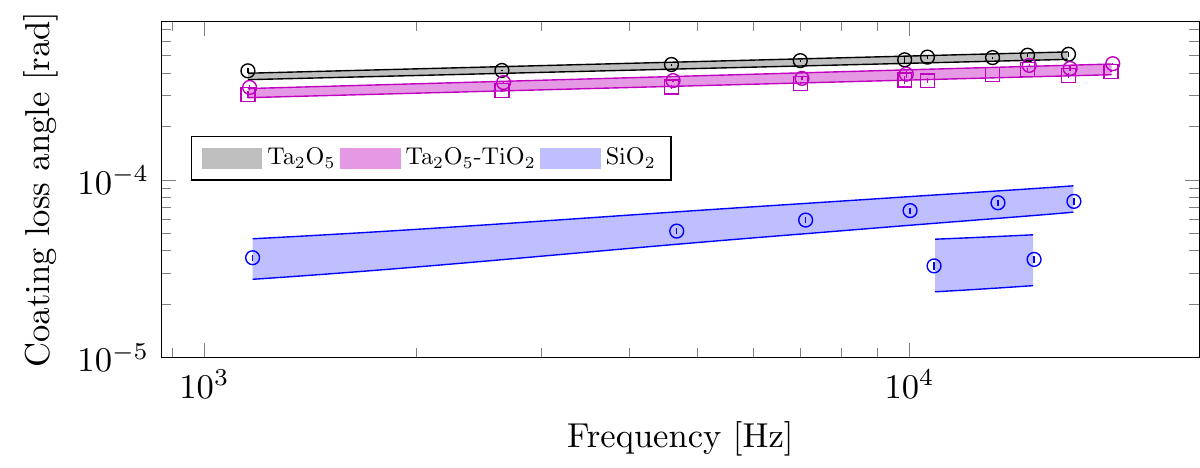}
	\caption{\label{FIG_stdMatGCann}Internal friction of coating materials of Advanced LIGO, Advanced Virgo and KAGRA: Ta$_2$O$_5$, Ta$_2$O$_5$-TiO$_2$, SiO$_2$. Error bars are shown, though barely visible; shaded regions represent uncertainties from fitting a frequency-dependent power-law loss model to each sample set: $\phi_c(f) = af^b$ for Ta$_2$O$_5$ and Ta$_2$O$_5$- TiO$_2$, $\phi_c(f) = af^b + \epsilon d \phi_e$ for SiO$_2$, where the $\epsilon d \phi_e$ term is the spurious contribution of the coated edge of the sample \cite{Cagnoli18}. The best-fit parameters of each data set are listed in Table \ref{TABLE_stdMatGCann}.}
\end{figure}

With a GeNS system, measurements of internal friction are performed via the ring-down method, i.e. by measuring the ring-down time of the vibrational modes of the sample: for the $i$-th mode of frequency $f_i$ and ring-down time $\tau_i$, the measured loss is $\phi_i = (\pi f_i \tau_i)^{-1}$, and the coating loss $\phi_{c\_i}$ can be written
\begin{equation}
	\phi_{c\_i} = [\phi_i + (D_i-1)\phi_{s\_i}]/D_i\ ,
\end{equation}
where $\phi_{s\_i}$ is the measured loss of the bare substrate. $D_i$ is the so-called dilution factor, defined as the ratio of the elastic energy of the coating, $E_c$, to the elastic energy of the coated disk, $E = E_c + E_s$, where $E_s$ is the elastic energy of the substrate; this ratio depends on the mode shape, and can be written as a function of the mode frequency and the mass of the sample before and after the coating deposition ($f_{s\_i}$, $f_i$, $m_s$ and $m$, respectively) \cite{Li14},
\begin{equation}
	D_i = 1 - \left(\frac{m_s}{m}\right)\left(\frac{f_{s\_i}}{f_i}\right)^2\ .
\end{equation}
Thus, thanks to the high reproducibility of frequency measurements allowed by a GeNS system, the dilution factor $D_i$ can now be measured; this implies that, unlike other experimental setups based on the ring-down method, our technique for loss characterization does not require prior knowledge of the coating Young's modulus and thickness (otherwise needed to estimate $D_i$). Furthermore, coating Young's modulus ($Y_c$) and Poisson's ratio ($\nu_c$) can be estimated by iteratively adjusting finite-element simulations of coated samples to fit the measured values of $D_i$: the best fit is obtained with a pair ($Y_c$, $\nu_c$) which minimizes the least-square figure of merit
\begin{equation}
	m_D = \sum_i \left(\frac{D_{\textrm{meas}\_i}-D_{\textrm{sim}\_i}}{\sigma_{\textrm{meas}\_i}}\right)^2\ ,
\end{equation}
where $D_{\textrm{sim}\_i}$ and $(D_{\textrm{meas}\_i} \pm \sigma_{\textrm{meas}\_i})$ are the simulated and measured dilution factors, respectively. Fig. \ref{FIG_stdMatGCann} compares the internal friction of coating materials of Advanced LIGO, Advanced Virgo and KAGRA mirrors, Table \ref{TABLE_stdMatGCann} summarizes their mechanical properties; all data have been measured with our GeNS system. We used a power-law model $\phi_c(f) = af^b$ to describe the observed frequency-dependent behavior of the coating internal friction \cite{Gilroy81,Travasso07,Cagnoli18}.
\begin{table}
	\centering	
	\caption{\label{TABLE_stdMatGCann}Mechanical properties of coating layers of Advanced LIGO, Advanced Virgo and KAGRA 
($a$ and $b$ are least-squares best-fit parameters of a power-law model $\phi_c(f) = af^b$ for the measured frequency-dependent internal friction shown on Fig. \ref{FIG_stdMatGCann}).}
	\begin{tabular}{lcccc}
	\hline 		
						& $a$ [10$^{-4}$ rad Hz$^{-b}$]	& $b$						& $Y_c$ [GPa]	& $\nu_c$\\
	\hline
	Ta$_2$O$_5$			& 1.88 $\pm$ 0.06				& 0.101 $\pm$ 0.004		& 117 $\pm$ 1	& 0.28 $\pm$ 0.01\\
 	Ta$_2$O$_5$-TiO$_2$ & 1.43 $\pm$ 0.07				& 0.109 $\pm$ 0.005		& 120 $\pm$ 4	& 0.29 $\pm$ 0.01\\
 	SiO$_2$ 			& 0.20 $\pm$ 0.04				& 0.030 $\pm$ 0.024		& 70 $\pm$ 1	& 0.19 $\pm$ 0.01\\
	\hline
	\end{tabular}
\end{table}

\begin{figure}
	\centering\includegraphics[width=\textwidth]{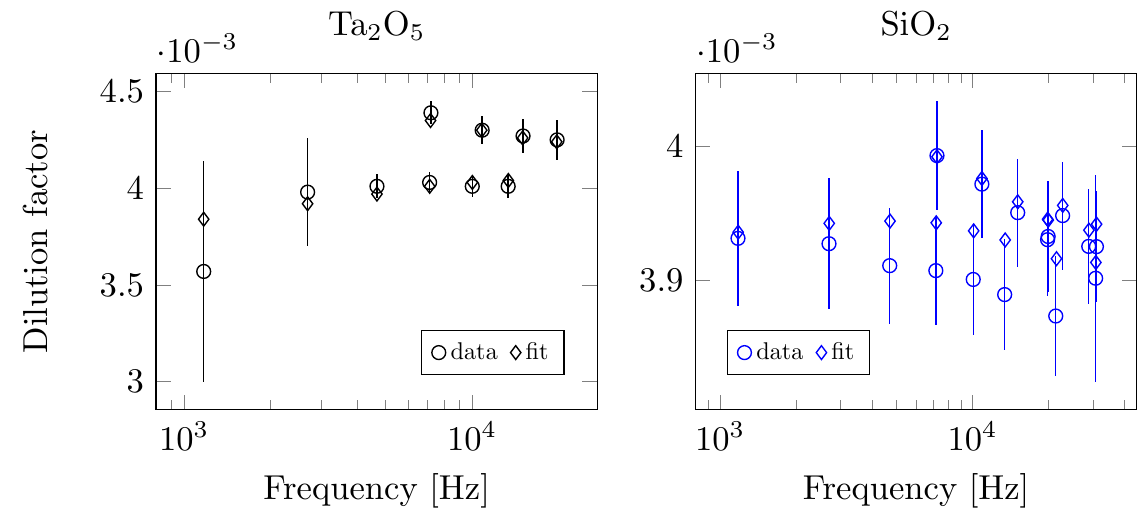}
	\caption{\label{FIG_dilFactCompar}Fit of measured dilution factors with simulated values, for Ta$_2$O$_5$ (left) and SiO$_2$ (right) coatings annealed 10 hours in air at 500 $^{\circ}$C. The scale of each plot is adapted to highlight the differences between measured and best-fit values.}
\end{figure}

Fig. \ref{FIG_dilFactCompar} shows the results of our fitting process for elastic constants of representative Ta$_2$O$_5$ and SiO$_2$ coating samples; the agreement between simulated and measured dilution factors vouches for the reliability of our method (though larger residuals for SiO$_2$ coatings seem to point out that either our model or our simulations might need further fine tuning in some specific case; this will be subject to further investigation).

\section{Reduction}
In a Fabry-Perot cavity like those of GW interferometers, the frequency-dependent CTN power spectral density can be written \cite{Harry02}
\begin{equation}
	S_{\textrm{CTN}} \propto \frac{k_BT}{2\pi f}\frac{t_c}{w^2}\phi_c\ ,
\end{equation}
where $f$ is the frequency, $T$ is the temperature, $t_c$ is the coating thickness, $\phi_c$ is the so called coating {\it loss angle} which quantifies the coating internal friction and $w$ is the laser beam radius. Thus, besides lowering the temperature of the mirrors \cite{Aso13,Hild11,Abbott17}, there are three key properties that may reduce CTN: coating thickness and internal friction, which depend on intrinsic properties of coating materials, and laser beam size, which requires the development of larger substrates and adapted deposition technology. Furthermore, the lowest coating thermal noise occurs when the coating Young's modulus is matched to that of the substrate \cite{Harry02}. Coating thickness is in turn a monotonically decreasing function of the refractive index contrast $c = n_{\textrm{\tiny{H}}}/n_{\textrm{\tiny{L}}}$ in the HR stack, where $n_{\textrm{\tiny{H}}}$ and $n_{\textrm{\tiny{L}}}$ are the high and low refractive indices, respectively; thus the larger $c$, the lower the coating thickness and hence the coating thermal noise (at constant reflection).

As a consequence, the optimal coating materials would feature the lowest internal friction and the largest index contrast at the same time, a Young's modulus matching as much as possible that of the substrate to be coated (73.2 GPa in fused-silica substrates used for room-temperature operation, $\geq 130$ GPa in silicon or sapphire substrates for cryogenic operation) and, in order to limit thermal lens effects, an optical absorption at least as low as it is to date, i.e. an extinction coefficient $k \sim 10^{-7}$ at $\lambda_0 = 1064$ nm (which is the wavelength of operation of GW interferometers) \cite{Pinard17}. By taking as a reference the optical \cite{Pinard17,Amato19} and mechanical (Table \ref{TABLE_stdMatGCann}) properties of the IBS oxide layers (Ta$_2$O$_5$, Ta$_2$O$_5$-TiO$_2$, SiO$_2$) within the HR coatings of Advanced LIGO, Advanced Virgo and KAGRA, future low-thermal-noise coatings should have $c > 1.44$ and $\phi_c < 10^{-4}$ at 100 Hz; these and other important coating requirements are summarized in Table \ref{TABLE_coatRequir}.
\begin{table}
	\centering	
	\caption{\label{TABLE_coatRequir}Tentative projection of requirements for optical and mechanical properties of coating layers for future GW interferometers, based on current standards (from \cite{Pinard17,Amato19}, this work).}
	\begin{tabular}{lc}
	\hline 		
	Refractive index		& $n_{\textrm{\tiny{H}}} > 2.09$\\
							& $n_{\textrm{\tiny{L}}} < 1.45$\\
	Extinction				& $10^{-7} < k < 10^{-6}$\\
	Scattering				& $\alpha_s \leq 10$ ppm\\
	Internal friction		& $\phi_c < 10^{-4}$ at 100 Hz\\
	Coated diameter			& $d \geq 35$ cm\\
	Thickness uniformity	& $\Delta t_c \leq 0.1$\%\\
	Surface roughness		& $\leq 0.1$ nm rms\\
	\hline
	\end{tabular}
\end{table}

In order to decrease CTN, several empirical options can be considered to further optimize the coatings of current GW interferometers: tuning of the sputtering ion beam, co-sputtering (also commonly referred to as {\it doping}), substrate heating during deposition and post-deposition annealing; in principle, these techniques could be also combined to cumulate their benefits. Also, alternative low-friction materials may be selected, then possibly further improved.

Finally, a more fundamental approach consists of isolating and possibly inhibiting the microscopic relaxation sites \cite{Gilroy81} which determine the amount of internal friction in coatings; this approach is based on the thorough investigation of the relation between the microscopic structure of coating materials and their macroscopic properties, both experimentally and through molecular dynamics simulations.

\subsection{Deposition parameters and annealing}
We studied the impact of different deposition parameters and of post-deposition annealing on the internal friction of silica (SiO$_2$) and tantala (Ta$_2$O$_5$) coatings, deposited with three different IBS coaters at LMA: the custom-developed so-called DIBS and Grand Coater (GC) and a commercial Veeco SPECTOR; the GC is used to coat the mirrors of GW detectors. Deposition parameters which may have a relevant impact on the coating properties are the beam ion energy and current and the geometric configuration of the elements inside the chamber, i.e. the distances and the angles between the sputtering sources, the sputtered targets and the substrates to be coated. Each coater has its own specific set of values for these parameters, optimized for yielding the highest coating optical quality, resulting in a different deposition rate. However, in all the coaters the ion energy and current are of the order of 1 keV and 0.1 A, respectively.
\begin{figure}
	\centering\includegraphics[width=\textwidth]{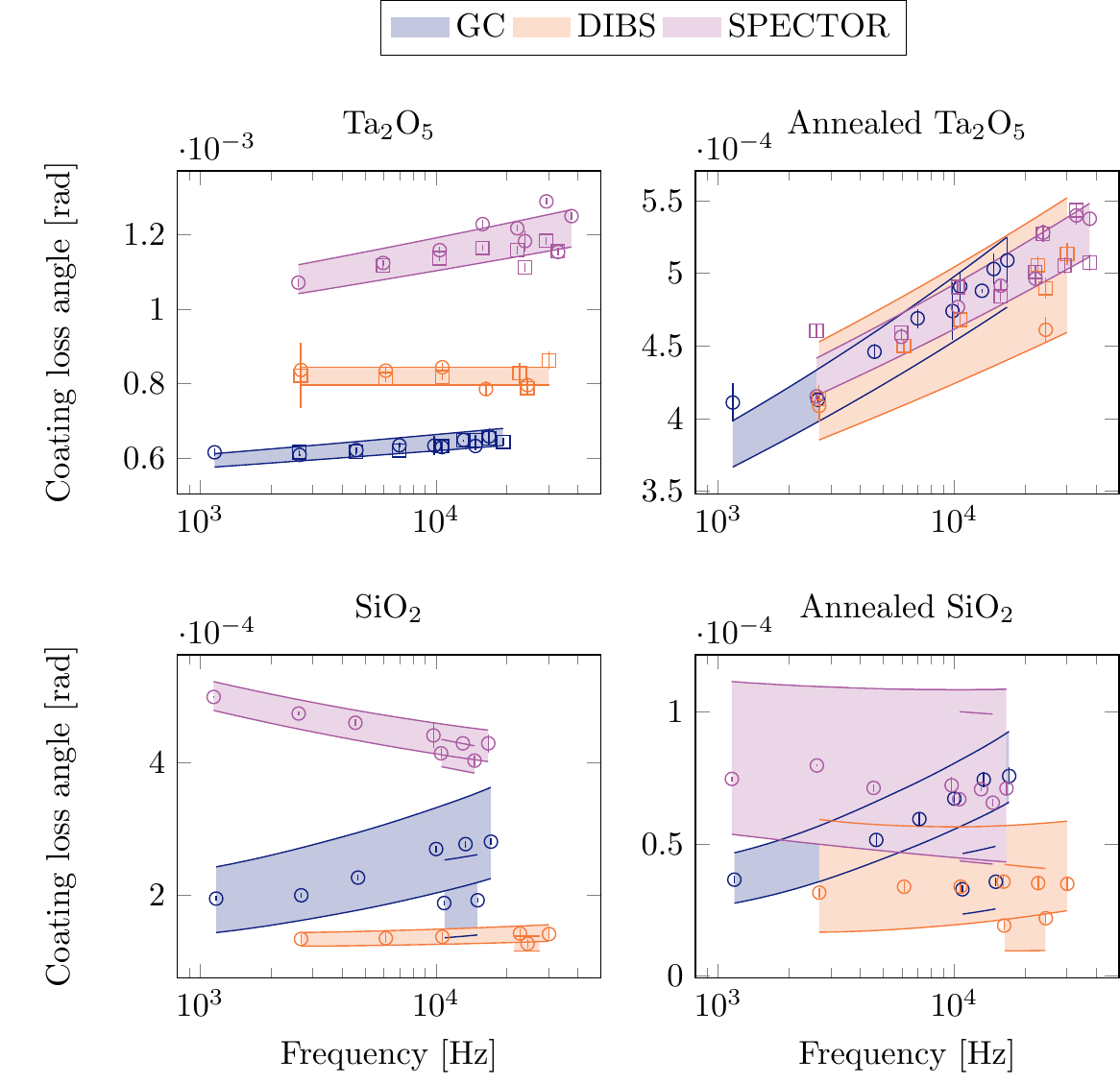}
	\caption{\label{FIG_Ta2O5_SiO2}Internal friction of Ta$_2$O$_5$ (top row) and SiO$_2$ (bottom row) coatings before (left column) and after (right column) 10-h in-air annealing at 500 $^{\circ}$C, deposited by different coaters: GC, DIBS and SPECTOR; shaded regions represent uncertainties from fitting a frequency-dependent power-law loss model to each sample set: $\phi_c(f) = af^b$ for Ta$_2$O$_5$ and $\phi_c(f) = af^b + \epsilon d \phi_e$ for SiO$_2$, where the $\epsilon d \phi_e$ term is the spurious contribution of the coated edge of the sample \cite{Cagnoli18}. The scale of each plot is adapted to highlight differences or similarities between different data sets.}
\end{figure}

Fig. \ref{FIG_Ta2O5_SiO2} shows the results of this preliminary study. We found no correlation between the measured loss and the distances of targets and substrates. For the Ta$_2$O$_5$ coatings, the GC provided the slowest rate and lowest loss values, whereas the SPECTOR the fastest rate (3 \r{A}/s) and the highest loss values. For the SiO$_2$ coatings, we observed that indeed the SPECTOR sample had the fastest deposition rate (2 \r{A}/s) and the highest loss values; however, despite having the same deposition rate (within 25\% experimental uncertainty), the DIBS sample had lower loss values than those of the GC sample. While this inconsistency will be subject to further investigation, we may conclude that, as a rule of thumb, the faster the deposition rate, the higher the loss.

The post-deposition annealing decreases the internal friction: depending on the initial values considered, the reduction of internal friction is of a factor 1.5 to 2.5 for Ta$_2$O$_5$ films, 5 to 6.5 for SiO$_2$ films. As the substrates are treated at 900 $^{\circ}$C prior deposition, the observed loss change upon annealing is due to the coating only. Once annealed, all the Ta$_2$O$_5$ films showed equal loss, as if their deposition history had been erased and an ultimate common structural configuration had been attained, whereas the gap between the loss values of SiO$_2$ films just decreased.

\subsection{Alternative oxides and doping}
\begin{figure}
	\centering\includegraphics[width=\textwidth]{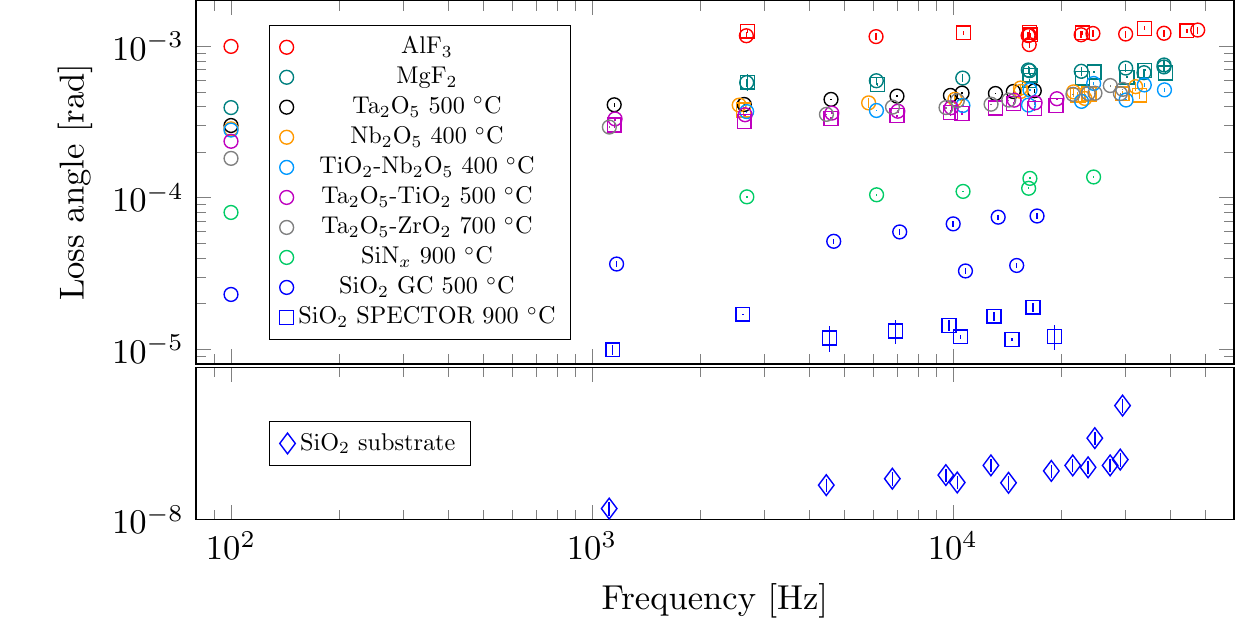}
	\caption{\label{FIG_coatLossCompar}Internal friction of IBS coatings (for annealed samples, the in-air annealing temperature $T_a$ is indicated in the legend and annealing time is 10 hours), compared to that of a representative substrate; values at 100 Hz are extrapolations from fitting a power-law loss model to each sample set.}
\end{figure}

To increase the refractive index contrast by replacing the current co-sputtered Ta$_2$O$_5$-TiO$_2$ layers with higher-index materials, we tested Ta$_2$O$_5$-TiO$_2$ coatings with different Ti/Ta mixing ratios \cite{Amato18}, Nb$_2$O$_5$ layers and co-sputtered TiO$_2$-Nb$_2$O$_5$ coatings. Also, following preliminary results from the LIGO Collaboration, we tried Ta$_2$O$_5$-ZrO$_2$ coatings treated at higher annealing temperatures. Fig. \ref{FIG_coatLossCompar} shows the results of internal friction for such oxide coatings; none of them yielded significantly lower friction, compared to the current Ta$_2$O$_5$-TiO$_2$ layers.

Interestingly, Fig. \ref{FIG_coatLossCompar} also shows that, despite their different nature and different annealing temperature $T_a$, all the high-index oxide coatings characterized so far (Ta$_2$O$_5$, Ta$_2$O$_5$-TiO$_2$, Nb$_2$O$_5$, TiO$_2$-Nb$_2$O$_5$ and Ta$_2$O$_5$-ZrO$_2$) have similar internal friction ($2 \cdot 10^{-4} \lesssim \phi_c \lesssim 5 \cdot 10^{-4}$) in the sampled frequency band, suggesting that this might be a general feature of this kind of coatings. Also, the measured values are in agreement with the behavior reported in the literature for many amorphous solids \cite{Topp96}.

\subsection{Fluorides}
\begin{figure}
	\centering\includegraphics[width=\textwidth]{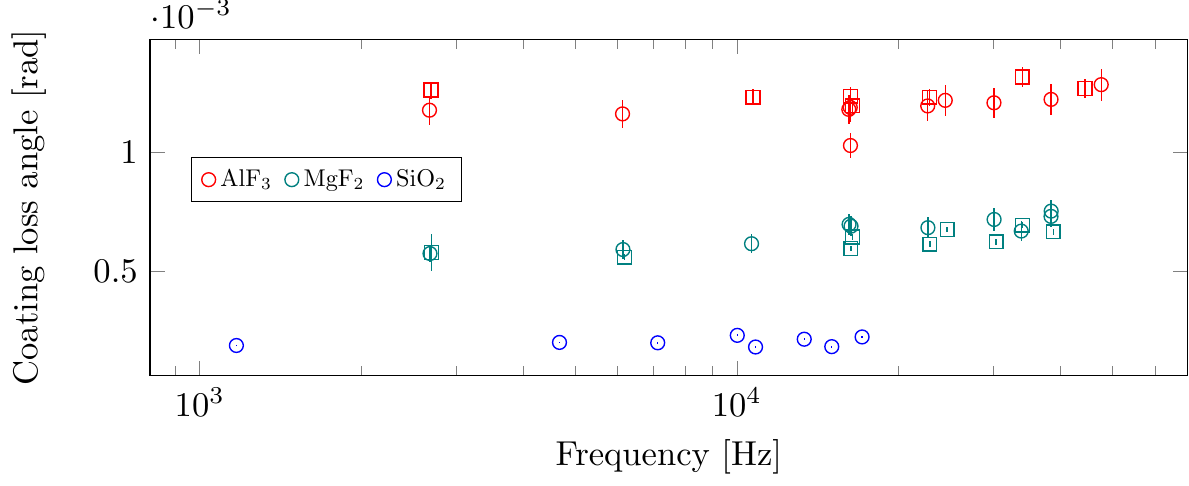}
	\caption{\label{FIG_fluorides}Internal friction of as-deposited IBS coatings: AlF$_3$ and MgF$_2$, compared to SiO$_2$ deposited in the GC.}
\end{figure}

To increase the refractive index contrast by replacing SiO$_2$ layers with lower-index materials, we characterized the optical properties and the internal friction of IBS MgF$_2$ and AlF$_3$ coatings produced by the {\it Laser Zentrum Hannover}. Before annealing, we measured a lower refractive index ($n_{\textrm{MgF\tiny{2}}} = 1.40$, $n_{\textrm{AlF\tiny{3}}} = 1.36$ at 1064 nm) but observed a too large optical absorption, $5 \cdot 10^{-5} < k < 10^{-4}$. Also, as Fig. \ref{FIG_fluorides} shows, their internal friction is much larger than that of current as-deposited SiO$_2$ layers. Work is ongoing to characterize the impact of annealing on their absorption and internal friction; soon we will also measure their low-temperature internal friction, for possible implementation in future interferometers \cite{Aso13,Hild11,Abbott17}.

\subsection{Silicon nitride}
\begin{figure}
	\centering\includegraphics[width=\textwidth]{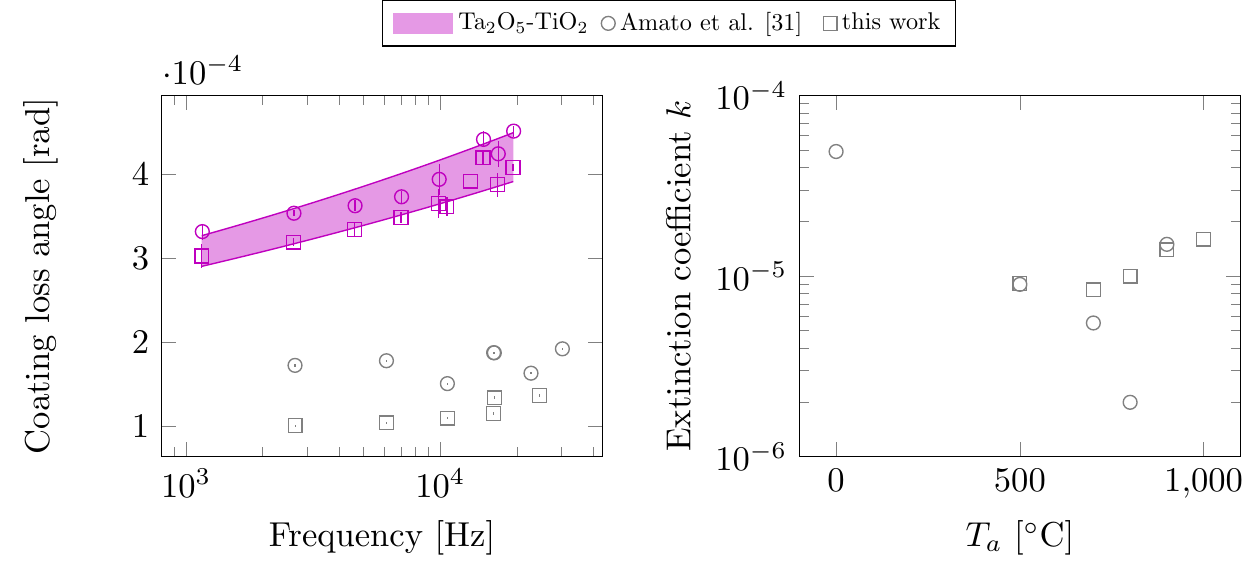}
	\caption{\label{FIG_SiNx}Characterization of IBS SiN$_x$ coatings. Left: internal friction of our first \cite{Amato18} and latest coating samples annealed in air at 900 $^{\circ}$C for 10 hours, compared to current values of Ta$_2$O$_5$-TiO$_2$ coatings annealed in air at 500 $^{\circ}$C for 10 hours (from Fig. \ref{FIG_stdMatGCann}). Right: extinction coefficient of our first \cite{Amato18} and latest coating samples as a function of the annealing temperature $T_a$ (in-air annealing time is 10 hours, $T_a = 0$ $^{\circ}$C corresponds to as-deposited coatings).}
\end{figure}

Because of their low internal friction \cite{Liu07}, we also decided to test silicon nitride (SiN$_x$) coatings. Usually, such coatings are deposited by chemical vapor deposition (CVD), suffering from high optical absorption ($k \sim 1-5 \cdot 10^{-5}$) due to hydrogen contamination \cite{Pan18}; moreover, their thickness uniformity still remains to be tested against the stringent requirements of GW detectors. Thus we chose to develop our own IBS SiN$_x$ coatings and test their loss versus the annealing temperature \cite{Amato18}. The latest results are shown in Figs. \ref{FIG_coatLossCompar} and \ref{FIG_SiNx}: after in-air annealing at 900 $^{\circ}$C for 10 hours, our IBS SiN$_x$ coatings have more than 3 times lower loss than present Ta$_2$O$_5$-TiO$_2$ coatings; however, though lower than that of CVD coatings, their absorption is still too high ($10^{-6} < k < 10^{-5}$), so we are presently working to further reduce this value.

Another remarkable advantage of silicon nitride coatings is their much higher crystallization temperature: we could anneal our IBS SiN$_x$ layers up to 900 $^{\circ}$C, without observing crystallization; since the crystallization temperature of our SiO$_2$ coatings is higher than 900 $^{\circ}$C \cite{Amato18}, an HR stack of SiN$_x$ and SiO$_2$ could be annealed at higher temperature, also decreasing the loss of SiO$_2$ layers \cite{Amato18} and thus the CTN of the whole stack itself.

\subsection{Amorphous silicon}
Recent results \cite{Liu14,Steinlechner18,Birney18} have shown that coatings of amorphous silicon (aSi) deposited through various techniques (e-beam evaporation, reactive low-voltage ion plating, IBS) can feature very high refractive index ($n \sim 3.6$) and very low internal friction ($10^{-5} < \phi_c < 10^{-4}$); however, their optical absorption ($10^{-5} < k < 10^{-4}$, for $1064 < \lambda < 2000$ nm and $20 < T < 290$ K) is still larger than the requirements of GW interferometers.

\subsection{Substrate heating}
Silicon nitride coatings prepared by low-pressure chemical-vapor deposition on substrates heated at 850 $^{\circ}$C \cite{Liu07} and aSi coatings deposited at relatively high temperatures ($T_s = 200-400$ $^{\circ}$C) by a variety of deposition techniques \cite{Liu14,Steinlechner18,Birney18} have shown that substrate heating during deposition is a promising technique to obtain coatings with very low internal friction ($10^{-6} < \phi_c < 10^{-4}$) at room and cryogenic temperature.

The high-temperature deposition of IBS Ta$_2$O$_5$ coatings has also been explored \cite{Vajente18}, but it seems to have negligible impact on the room-temperature internal friction of coatings deposited at $150 < T_s < 500$ $^{\circ}$C.

At LMA, we have now completed the installation of a rotating heating substrate holder, to deposit uniform single layers and HR coatings up to $T_s = 800$ $^{\circ}$C. Soon we will use it to test the deposition of Ta$_2$O$_5$ coatings at temperatures right below their crystallization limit ($500 < T_s < 650$ $^{\circ}$C) \cite{Amato18} and of SiO$_2$ and silicon nitride coatings, whose crystallization occurs beyond 900 $^{\circ}$C \cite{Amato18}.

\subsection{Composite high-index layers}
In order to replace the current Ta$_2$O$_5$-TiO$_2$ high-index layers of GW interferometers, the use of stacks of TiO$_2$ and SiO$_2$ with nm-thick layers has been proposed \cite{Pan14,Magnozzi18,Kuo19}. Though they have a lower refractive index ($n = 1.76$) \cite{Kuo19}, these nm-layered stacks can be annealed at higher temperature ($700 < T_a < 800$ $^{\circ}$C) and have low internal friction at temperatures below 100 K \cite{Kuo19}. However, to date, the deposition of a full HR stack embedding composite layers is yet to be achieved.

\subsection{Multi-material HR stacks}
In order to conjugate low internal friction and low optical absorption, HR stacks composed of at least three different coating materials have been proposed \cite{Steinlechner15,Yam15}. In this multi-material design (like, for instance, in Ta$_2$O$_5$/SiO$_2$/aSi \cite{Steinlechner15}, Ta$_2$O$_5$-TiO$_2$/SiO$_2$/SiN$_x$ \cite{Pan18b} and Ta$_2$O$_5$/SiO$_2$/HfO$_2$-SiO$_2$/aSi \cite{Craig19} stacks), the low-friction, absorptive layers are buried under the low-absorption, dissipative layers.

Multi-material HR stacks might be a viable solution to substantially decrease CTN and, at the same time, to fulfill the optical requirements of future GW detectors (of cryogenic ones in particular \cite{Pan18b,Craig19}). However, to date, an experimental demonstration of such systems is yet to be achieved.

\subsection{Crystalline coatings}
Mono-crystalline semiconductors grown by molecular-beam epitaxy like GaAs/AlGaAs \cite{Cole13} and GaP/AlGaP \cite{Murray17} coatings feature low internal friction ($\phi_c \leq 10^{-4}$) at ambient and cryogenic temperature. Furthermore, GaAs/AlGaAs coatings have competitive optical properties \cite{Cole16} when compared with current state-of-the art IBS coatings \cite{Pinard17}. However, the implementation of GaAs/AlGaAs coatings in high-finesse cavities presents issues yet to be solved: i) they are not available in the large diameter ($d \geq 35$ cm) required by GW interferometers, as their size is limited to date by the available size of the lattice-matched GaAs wafers required for seeded growth; ii) the presence of a number of imperfections \cite{Penn19}, including point defects $>50$ $\mu$m in diameter and un-bonded regions (due to the transfer of the mono-crystalline stack from the growth wafer to the target optical surface) as well; iii) the inconsistency between the latest measured values \cite{Penn19} and previous estimations \cite{Cole13} of coating internal friction.

The GaP/AlGaP coatings have been specifically designed to overcome the current size limitation of GaAs/AlGaAs-based systems. However, due to the limited index contrast between GaP and AlGaP layers, the number of layers $N$ required by a GaP/AlGaP design to achieve the same optical reflectivity of current gravitational-wave detectors' HR coatings is large, $N>100$ \cite{Cumming15}; to date, $N$ is limited to considerably lower values \cite{Cumming15}, for a limited reflectivity.

\subsection{Structural analyses and models}
In the commonly accepted phenomenological two-level systems model \cite{Gilroy81}, for temperatures of more than several Kelvin, internal friction arises from thermally-activated transitions of particle structures between equilibrium configurations within the potential energy landscape of an amorphous solid. Indeed, the TLS model still does not comprehend several crucial details of the energy loss, such as the identification of the structure relaxing from one equilibrium position to another and the factors that determine a particular energy distribution of TLS, for instance. As a result, a considerable amount of theoretical work is presently ongoing in order to overcome the TLS model.

As a matter of fact, the utmost scientific and technological challenge of CTN reduction may lie in the possibility of isolating the relaxation sites, whose nature has remained elusive to date: by understanding which properties of amorphous structures are able to affect either the rate or the number of such transitions, it could be possible to inhibit relaxation and hence internal friction.

In recent years, the search for a structural origin of internal friction in amorphous materials has produced several sound results, from experimental correlations between dissipation and structural organization \cite{Bassiri13,Granata18,Amato19b} to molecular dynamics simulations \cite{Hamdan14,Trinastic16,Prasai19,Puosi19}.

\section{Conclusions}
Coatings of precision optical and quantum transducers are required to simultaneously feature outstanding optical and mechanical properties: low internal friction for low thermal noise, low optical absorption and scattering for low optical loss (also, for km-scale interferometric gravitational-wave detection, extreme surface figure requirements). These specifications are extraordinarily difficult to meet in a single coating material and deposition technique; to date, some promising options are IBS SiN$_x$ and SiO$_2$ \cite{Amato18} and amorphous silicon \cite{Liu14,Steinlechner18,Birney18}, to be deposited \cite{Liu07,Liu14} and possibly further annealed \cite{Amato18,Liu07} at high temperature. Such materials, if used in a multi-material design \cite{Steinlechner15,Yam15}, might be a viable solution for the next sensitivity upgrades of Advanced LIGO and Advanced Virgo (2022-2023) and KAGRA.

In the meanwhile, structural analyses \cite{Bassiri13,Granata18,Amato19b} and molecular dynamics simulations \cite{Hamdan14,Trinastic16,Prasai19,Puosi19} might provide critical hints towards a more radical solution of the coating thermal noise issue, to benefit future GW interferometers \cite{Hild11,Abbott17}. 

\section*{Acknowledgments}
Portions of this work were presented at the Optical Society of America (OSA) Conference on Optical Interference Coatings (OIC) in 2019, paper FA.1.

\section*{Disclosures}
The authors declare that there are no conflicts of interest related to this article.

\end{document}